\documentclass[11pt,a4paper]{article}
\usepackage{jheppub}
\usepackage[T1]{fontenc}
\usepackage{amsmath}
\usepackage{amssymb}
\usepackage{esint}
\usepackage{amsfonts}
\usepackage{graphicx,color}
\usepackage{slashed}
\usepackage{young}
\usepackage[vcentermath]{youngtab}
\usepackage[utf8]{inputenc}
\usepackage{tensor}
\usepackage{etoolbox}
\usepackage{bbm}
\usepackage{soul}
\usepackage[normalem]{ulem}

\makeatletter

\newcommand{\be}{\begin{equation}}
\newcommand{\ee}{\end{equation}}
\newcommand{\bea}{\begin{eqnarray}}
\newcommand{\eea}{\end{eqnarray}}

\renewcommand{\hat}{\widehat}

\renewcommand{\d}{\partial}

\def\cL{\mathcal{L}}

\newcommand*\xbar[1]{%
  \hbox{%
    \vbox{%
      \hrule height 0.5pt 
      \kern0.3ex
      \hbox{%
        \kern-0.0em
        \ensuremath{#1}%
        \kern-0.0em
      }%
    }%
  }%
} 

\pdfpageheight\paperheight
\pdfpagewidth\paperwidth

\pdfoutput=1 

    \patchcmd{\maketitle}{\@fpheader}{}{}{}

\title{Hamiltonian structure and asymptotic symmetries of the Einstein-Maxwell system at spatial infinity}
\author[a,b]{Marc Henneaux,}
\author[c]{and C\'edricTroessaert}
\affiliation[a]{Universit\'e Libre de Bruxelles and International Solvay Institutes, ULB-Campus Plaine CP231, B-1050 Brussels, Belgium}
\affiliation[b]{Coll\`ege de France, 11 place Marcelin Berthelot, 75005 Paris, France}
\affiliation[c]{Max-Planck-Institut f\"{u}r Gravitationsphysik (Albert-Einstein-Institut),
Am M\"{u}hlenberg 1, \\ DE-14476 Potsdam, Germany}

\abstract
{We present a new set of asymptotic conditions for gravity at spatial infinity that includes gravitational magnetic-type solutions, allows for a non-trivial Hamiltonian action of the complete $BMS_4$ algebra, and leads to a non-divergent behaviour of the Weyl tensor as one approaches null infinity.  We then extend the  analysis to the coupled Einstein-Maxwell system and obtain as canonically realized asymptotic symmetry algebra a semi-direct sum of the $BMS_4$ algebra with the angle dependent $u(1)$ transformations. The Hamiltonian charge-generator associated with each asymptotic symmetry element is explicitly written. The connection with matching conditions at null infinity is also discussed.}

\makeatother

\begin{document}
\maketitle \flushbottom

\section{Introduction}
\setcounter{equation}{0}

The BMS symmetry was originally discovered at null infinity in the context of gravitational radiation in asymptotically flat spacetimes \cite{Bondi:1962px,Sachs:1962wk,Sachs:1962zza,Penrose:1962ij,Madler:2016xju,Alessio:2017lps}.  A major development in the field has been the realization that soft graviton theorems could be interpreted as Ward identities for the BMS asymptotic symmetries (for a review, see \cite{Strominger:2017zoo}).   Now, Ward identities can only be derived from bona fide conserved charges.  This raises the question of constructing the  BMS charges that canonically generate the BMS symmetries.

The question is not entirely trivial since it has been long appreciated that at null infinity, the natural concepts to be considered are fluxes, rather than charges, which are not conserved whenever the fluxes are non zero \cite{Dray:1984rfa,Wald:1999wa,Barnich:2011mi,Bunster:2018yjr}.  The symmetries are in fact not even canonically generated and the association of functions with symmetries is therefore intricate.  The situation is somewhat analogous to the dynamics of a system in a box with semi-permeable boundary conditions that allow non-vanishing outgoing (or incoming) fluxes. Hypersurfaces that ``reach'' null infinity are non-Cauchy. As one moves from one non-Cauchy hypersurface to the next, the past (or future) development shrinks with information leaking to (or coming from) null infinity.  It is only when the fluxes at null infinity (known as "the news" in the gravitational case) vanish that one recovers a standard Hamiltonian picture.

By contrast, the description of the dynamics on Cauchy hypersurfaces  is Hamiltonian even if there is gravitational radiation, since Cauchy hypersurfaces capture the whole dynamical system.  There is no flux at spatial infinity and any symmetry is directly generated by a conserved charge that can be determined by standard canonical techniques, without having to impose the physically rather restrictive condition that the news is zero. The charge-generators for the full BMS group, including supertranslations,  can thus be worked out in principle at spatial infinity, by considering the dynamical variables on Cauchy hypersurfaces.

Until recently, however, Hamiltonian analyses at spatial infinity failed to exhibit the $BMS_4$ algebra  as a genuine asymptotic symmetry with well-defined charges.  Either the boundary conditions were invariant under a bigger infinite-dimensional algebra, but the elements of that algebra had generically divergent charges; or the boundary conditions, taken to be more restrictive to avoid these divergences,  admitted then as non-trivial canonical asymptotic symmetries only the finite-dimensional Poincar\'e algebra.

In order to resolve this somewhat schizophrenic tension between null infinity and spatial infinity, a new set of boundary conditions were given in \cite{Henneaux:2018cst} for pure gravity at spatial infinity. These new conditions were shown to be invariant under $BMS_4$ supertranslations, which acted non-trivially.
But they had two unsatisfactory features.  The first is that they excluded solutions with non-zero gravitational magnetic charges, such as the Taub-NUT metric. The second was the presence of solutions that developed logarithmic divergences at null infinity and broke the differentiability conditions usually accepted there (while remaining finite at spatial infinity) \cite{Troessaert:2017jcm}. 

A similar difficulty was pointed out in \cite{Henneaux:2018gfi} for electromagnetism and solved there.  Following the lines of that article (in which the results of the present note were in fact announced), we give here a new set of boundary conditions for gravity at spatial infinity that keeps the good properties of the boundary conditions of \cite{Henneaux:2018cst} while avoiding their annoying features.   To be specific, these boundary conditions are such that
\begin{enumerate}
	\item  Solutions with gravitational magnetic mass belong to the phase-space so defined (in addition to  solutions with gravitational electric mass such as Schwarzschild or Kerr).
	\item The symplectic form, and thus the kinetic term in the action, is finite.
	\item The asymptotic symmetry is the $BMS_4$ algebra, all elements of which have  well-defined canonical generators.
\end{enumerate}

We then extend the analysis to the coupled Einstein-Maxwell system.  We also compute the Poisson brackets of the charges and verify that their algebra is the natural semi-direct product of the $BMS_4$ algebra with the abelian algebra of angle-dependent $u(1)$ transformations\footnote{The charges are given by surface contributions plus  weakly vanishing bulk terms.   The bulk terms are arbitrary but can be determined by means of gauge conditions, corresponding to the fact that global symmetries are determined up to (proper) gauge transformations in the absence of gauge fixing.  Once gauge conditions are imposed, one must use the corresponding Dirac bracket  instead of the Poisson bracket.  For gauge-invariant observables, however, the two coincide (up to weakly vanishing terms set anyway strongly equal to zero) -- and coincide also with the reduced phase space Poisson bracket.   For that reason, it has become customary not to make the distinction between Poisson and Dirac brackets in this context.}.

Our paper is organized as follows. We first consider pure gravity. In Section \ref{sec:philosophy}, we explain the general idea underlying the choice of boundary conditions, starting from the pioneering work \cite{Regge:1974zd} where parity conditions were imposed asymptotically.    We then give in Section \ref{sec:grav} the precise form of these boundary conditions.  To streamline the presentation, we first give their explicit form, which turns out to be unexpectedly simple, and verify next that they possess all the good properties that they should have. In Section \ref{sec:EM}, we review the
asymptotic structure of free electromagnetism in Minkowski space and then combine it
with the gravitational one in order to describe the asymptotic structure of
the Einstein-Maxwell system. Section \ref{sec:Conclusions} is devoted to conclusions and comments, including a discussion of the connection with other works. In
Appendix \ref{app}, we give more details on the structure of the new asymptotic
conditions for gravity.

We close this introductory section with an important question.  Why can one state that the $BMS_4$ symmetries exhibited at spatial infinity are the same $BMS_4$ symmetries found at null infinity?  This question was answered already in \cite{Troessaert:2017jcm,Henneaux:2018cst,Henneaux:2018gfi}, but since this is an important point, we examine it again here from a somewhat more conceptual standpoint.   

A symmetry is a transformation that leaves the action invariant.  Symmetries are conveniently discussed in the Hamiltonian formulation, where the action is $S[q^i(t), p_i(t)] = \int dt \left(p_i \dot{q}^i - H\right)$.  The Hamiltonian can have an explicit time dependence.  A symmetry transformation must preserve the kinetic term of the action and so must be a canonical transformation.  Let $G$ be its generator.  It is a phase space function that may depend explicitly on $t$.  The action will be completely invariant (up to a term at the time boundaries) if and only if $\frac{dG}{dt} \equiv \frac{\partial G}{\partial t} + [G, H] = 0$. This equation shows that giving a symmetry transformation at a time $t_0$, say, (i.e., its generator $G(t_0)$), determines it uniquely at all times. The same is true in field theory: giving a symmetry on a Cauchy hypersurface determines it everywhere (modulo proper gauge transformations \cite{Benguria:1976in} in the case of gauge theories).  

The same is also true for asymptotic symmetries where the relevant dynamics is the asymptotic dynamics, which is fixed once one has chosen the time slicing at infinity.  In the case relevant to our question, in order to do the matching between spatial infinity and null infinity, one must integrate the equation for the symmetry generator from the given Cauchy hypersurface to the ``critical spheres'', i.e., the past of future null infinity and the future of past null infinity.  To that end, it is convenient to blow up spatial infinity, which is a single point in Penrose's compactification \cite{Penrose:1962ij}, to a cylinder bounded by the critical spheres, and to adopt a time slicing that covers that cylinder, such as hyperbolic coordinates  \cite{Ashtekar:1978zz,BeigSchmidt,Beig:1983sw,Ashtekar:1991vb,Compere:2011ve} or the coordinates introduced in \cite{Fried1,Friedrich:1999wk,Friedrich:1999ax}.  In practice, since any transformation that leaves the action invariant also maps solutions on solutions, one can determine the asymptotic symmetry from its ``initial data'' by requesting that the equations of motion at infinity be preserved.  This is the method followed in  \cite{Troessaert:2017jcm,Henneaux:2018cst,Henneaux:2018gfi} for pure gravity and the Einstein-Maxwell system, leading to a precise (and somewhat subtle) matching between the descriptions of the $BMS_4$ symmetry at spatial infinity and at null infinity.

\section{Parity conditions}
\label{sec:philosophy}

Our boundary conditions strongly rely on the  approach developed in \cite{Regge:1974zd},  where parity conditions were imposed on the leading orders of the asymptotic fields.  

The rationale behind the parity conditions can be understood as follows.  If one considers the usual asymptotics for gravity without parity conditions, the bulk integral of the symplectic structure acquires a logarithmic divergence for generic configurations. Now, a finite symplectic form is necessary for having a well-defined canonical structure. In order to cancel the divergence, one must therefore restrict further the asymptotic fields.  This can be done by means of parity conditions. 

The natural choice for these parity conditions \cite{Regge:1974zd} is  such that the resulting symmetry algebra reduces to the exact symmetries of the background, i.e., the Poincar\'e algebra.  This is because any supertranslation which is not a translation is either trivial -- i.e., is pure gauge with zero charge -- or forbidden -- i.e.,  does not preserve the parity conditions.   A different choice of parity conditions must thus be considered.

The different boundary conditions proposed in  \cite{Henneaux:2018cst} involve a twist in the parity conditions for the leading orders of the angular components of the metric. These new conditions, while still being Lorentz invariant, keep the symplectic structure finite and allow for the non-trivial enhancement of the asymptotic symmetry algebra to the $BMS_4$ algebra.  However, as we pointed out above, the twist implies that solutions with non-zero gravitational magnetic charges, like the Taub-NUT metric, are not part of the phase-space.  The way out is suggested by electromagnetism. 

Electromagnetism enlarges  further the symmetry by introducing asymptotic angle-dependent $u(1)$ transformations, which were identified at null infinity in \cite{Strominger:2013lka,Barnich:2013sxa}.  A similar  picture holds, namely, that the angle-dependent $u(1)$ transformations are not canonically generated at null infinity whenever the ``electromagnetic news function'' does not vanish.  In order to exhibit the conserved charge-generators, the precise asymptotic structure of electromagnetism at spatial infinity  (i.e.,  on Cauchy hypersurfaces) has been studied  in \cite{Henneaux:2018gfi} (earlier important work include \cite{Balachandran:2013wsa}).   Again, the symplectic structure diverges unless one imposes extra conditions which can be  parity conditions. The natural parity conditions, given in \cite{Henneaux:1999ct}, suffers however from the same drawback as the corresponding ones for gravity: they freeze the possibility to perform non trivial angle-dependent  $u(1)$ transformations, except constant ones.  A twist of the parity conditions on the angular components of the asymptotic fields is thus also necessary in this case if one wants a non-trivial action of the angular dependent $u(1)$-transformations present at null infinity.  But as in the gravitational case, while imposing these twisted parity conditions leads to a well-defined system, this choice excludes magnetic monopoles and introduces solutions possessing a logarithmic divergent electromagnetic field at null infinity.  

However, it was shown in \cite{Henneaux:2018gfi} that it is sufficient for this twist to be an improper gauge transformation (for ``improper'', we follow the terminology of \cite{Benguria:1976in}). This leads to the introduction of a set of hybrid parity conditions combining the best of both choices: a non-trivial action of the enhanced asymptotic symmetry algebra, the absence of solutions diverging at null infinity and the possibility to describe magnetic monopoles.  

We show in this paper that the same procedure also works for gravity.  The
gravitational analog of the hybrid parity conditions introduced for
electromagnetism are a mixture of the parity conditions of \cite{Regge:1974zd}
with those of \cite{Henneaux:2018cst}.  More precisely, the twisted parity
component of \cite{Henneaux:2018cst} that we include here is an improper gauge transformation (i.e., diffeomorphism). These hybrid parity conditions for gravity form a well-defined system and allow for a non-trivial action of the BMS4 supertranslations without the drawbacks mentioned earlier.   They are given in the next section.

The new boundary conditions (``parity of \cite{Regge:1974zd} on the asymptotic fields modulo an arbitrary improper gauge transformation'') are deceptively simple.  While they are straightforwardly invariant under $BMS_4$ transformations, which are improper gauge transformations, they lead to complications in the verification of the finiteness of the symplectic form and of the charges, which holds only if additional conditions are imposed at infinity.  These extra conditions are also explicitly spelled out.  In the case of electromagnetism, the twist of the parity conditions by an improper gauge transformation also leads to the surprising feature that an extra surface degree of freedom must be introduced in order for the generators of Lorentz transformations to exist \cite{Henneaux:2018gfi}.  No such additional degree of freedom is necessary in the case of gravity.

\section{Pure gravity}
\label{sec:grav}

As announced in the introduction, we first give the boundary conditions and then check that they have the requested properties. 

The hamiltonian action of gravity can be written as
\begin{gather}
	S  =\int dt \left\{ \int d^3x \left( \pi^{ij} \d_t g_{ij}  - N^i \mathcal
		H^{grav}_i - N \mathcal H^{grav} \right) - B_{S_\infty}  \right\},\\
	\mathcal H^{grav}  = - \sqrt g R + \frac{1}{\sqrt g} (\pi^{ij} \pi_{ij} -
	\frac{1}{2} \pi^2),\quad
	\mathcal H^{grav}_i  = -2 \nabla_j \pi^j_i.
\end{gather}
where $B_{S_\infty}$ is a boundary term on the sphere at spatial infinity that depends on the asymptotic values of the lapse and the shift (see below).

\subsection{Boundary conditions} 
The boundary conditions on the dynamical variables that define asymptotically flat space-times are, in spherical coordinates:
\begin{equation}
	\label{eq:asymptgravI}
	\begin{array}{rclrcl}
		g_{rr} &=& 1 + \frac{1}{r} \xbar h_{rr} + \frac{1}{r^2} h^{(2)}_{rr} +
		o(r^{-2}),&
		\pi^{rr} &=& \xbar \pi^{rr} + \frac{1}{r} \pi^{(2)rr} +
		o(r^{-1}),\\
		g_{rA} &= &\frac{1}{r} h^{(2)}_{rA} +
		o(r^{-1}),&
		\pi^{rA} & = & \frac{1}{r} \xbar \pi^{rA} + \frac{1}{r^2} \pi^{(2)rA} +
	o(r^{-2}),\\
	g_{AB} & = & r^2 \xbar\gamma_{AB} + r \xbar h_{AB} +  h^{(2)}_{AB} +
	o(1),& 
	\pi^{AB} & = & \frac{1}{r^2} \xbar \pi^{AB} + \frac{1}{r^3} \pi^{(2)AB}+
	o(r^{-3}).
\end{array}
\end{equation}
The asymptotic $2$-dimensional metric $\xbar \gamma_{AB}$ is  here the usual metric on the sphere.   In Cartesian coordinates, the decay expressed by these conditions is the standard one, namely, $g_{ij} - \delta_{ij} \sim \frac{1}{r}$ and $\pi^{ij} \sim \frac{1}{r^2}$.

The leading orders of the dynamical variables are further subject to two types of extra conditions:
(i) Parity conditions; (ii)  Constraint conditions.

In order to describe these conditions,  it is useful to introduce a 1+2 radial split of the 3 dimensional metric $g_{ij}$:
\begin{gather}
	g_{rr} = \lambda^2 +  \gamma_{AB}\lambda^A \lambda^B, \quad g_{rA} =
	\gamma_{AB}\lambda^B, \quad g_{AB} = \gamma_{AB}, \\
	\lambda = 1 + r^{-1} \xbar \lambda +  r^{-2} \lambda^{(2)} +
	o(r^{-2}), \quad \lambda^A = r^{-3} \lambda^{(2)A} + o(r^{-3}), \\
	\gamma_{AB} = r^2 \xbar \gamma_{AB} + r \xbar h_{AB} + h^{(2)}_{AB} +
	o(1).
\end{gather}
We will use $D_A$ and $\xbar D_A$ to respectively denote the covariant derivatives of $\gamma_{AB}$ and $\xbar \gamma_{AB}$. The indices $A, B, ...$ on bulk fields will be lowered and raised with $\gamma_{AB}$ and its inverse $\gamma^{AB}$ while the same indices on asymptotic fields will be lowered and raised with $\xbar\gamma_{AB}$ and its inverse $\xbar \gamma^{AB}$.  The extrinsic curvature of the constant $r$ surfaces is then given by
\begin{equation}
	K_{AB} = \frac{1}{2\lambda} (-\d_r \gamma_{AB} + D_A \lambda_B + D_B
	\lambda_A), \quad K^A_B = -r^{-1} \delta^A_B + r^{-2} \xbar k^A_B +
	r^{-3} {k^{(2)}}^A_B + o(r^{-3}).
\end{equation}

\subsubsection*{Constraint conditions}

The constraints are requested to asymptotically decay faster than what (\ref{eq:asymptgravI}) implies, i.e., one imposes: 
\begin{equation}
	\label{eq:asymptconstgrav}
	\mathcal H^{grav} = o(r^{-1}),\quad \mathcal H^{grav}_r = o(r^{-1}),\quad \mathcal
	H^{grav}_A = o(1).
\end{equation}
In terms of the asymptotic fields, these conditions are:
\begin{gather}
	\label{eq:asymptconst}
	\xbar D_A \xbar D_B \xbar k^{AB} -
	\xbar D_A \xbar D^A \xbar k = 0, \quad \xbar \pi^{rA} +\xbar D_B \xbar
	\pi^{AB} = 0,\quad \xbar D_A \xbar D_B \xbar
	\pi^{AB} + \xbar \pi^A_A = 0.
\end{gather}

\subsubsection*{Parity conditions}

We further request that the leading part of the asymptotic fields fulfill the following parity conditions under the sphere antipodal map, in coordinates where this map takes the form\footnote{Note that in terms of standard spherical coordinates, the antipodal map is actually $\theta \rightarrow \pi - \theta$ and $\varphi \rightarrow \varphi + \pi$ (and $r \rightarrow r$).  This implies $d \theta \rightarrow - d \theta$ and $d \varphi \rightarrow d \varphi$.  Therefore, the condition that $k_{AB}$ is even (for example), i.e., $k_{AB} (-x^B) =  k_{AB}(x^B)$, is equivalent to the statement that $k_{\theta\theta}$ and $k_{\varphi\varphi}$ are even and $k_{\theta\varphi}$ is odd. \label{foot:Parity}} $x^A \rightarrow -x^A$:
\begin{gather}
	\label{eq:parityuptogI}
	\xbar \lambda = \text{even}, \quad \xbar \pi^{rr} - \xbar \pi^A_A =
	\text{odd}, \quad \xbar \pi^{rA} =
	(\pi^{rA})^{even} - \sqrt{\xbar\gamma}\xbar D^A V, \\
	\xbar \pi^{AB} = (\xbar \pi^{AB})^{odd} + \sqrt{\xbar\gamma}(\xbar D^A \xbar D^B V- \xbar
	\gamma^{AB} \xbar D_C \xbar D^CV), \\ \xbar k_{AB} =
	(\xbar k_{AB})^{even} + \xbar D_A \xbar D_B U + U \xbar \gamma_{AB}.
	\label{eq:parityuptogII}
\end{gather}
Here, it is the {\em same} scalar $V$ that appears in the parity conditions of $\xbar\pi^{rA}$ and $\xbar \pi^{AB}$. One easily sees that the two functions $U$ and $V$ can be restricted to be odd and even, respectively, since the contributions of the opposite parity terms can be absorbed through a redefinition of $(\pi^{rA})^{even}$, $(\xbar \pi^{AB})^{odd}$ and $(\xbar k_{AB})^{even}$.

\subsubsection*{Comparison with previous parity conditions}

The asymptotic decay (\ref{eq:asymptgravI}) with the above constraint and parity conditions define our phase space.  
In Appendix \ref{app}, Subsection \ref{app:weyl}, we rewrite these boundary conditions more invariantly in terms of the components of the asymptotic Weyl tensor, which are shown to possess definite parities.  

If one sets $U=V=0$ in the parity conditions, one recovers the parity conditions of \cite{Regge:1974zd}.  This shows that the phase space defined by the boundary conditions includes the Schwarzschild and Kerr metrics as well as their transformed under asymptotic boosts.  It also includes the Taub-NUT solution \cite{Bunster:2006rt} but one does not need to impose the symmetric gauge conditions adopted asymptotically in that reference to see it.  

If one sets $(\xbar \pi^{rA})^{even}$, $(\xbar \pi^{AB})^{odd}$ and $(\xbar k_{AB})^{even}$ to zero, one gets a subset of the configurations considered in \cite{Henneaux:2018cst}, which have these fields of opposite parity, i.e., $ \xbar \pi^{rA} = \text{odd}$, $\xbar \pi^{AB} = \text{even}$ and $\xbar k_{AB} = \text{odd}$ (``twisted parity conditions'').  We do include here angular components of twisted parity, but they have to take the specific ``improper gauge'' form parametrized by the functions $U$ and $V$.  It is only under this  specific form that twisted parity components are compatible with generic components of untwisted parity,  allowing thereby the Taub-NUT metric.

\subsection*{Lapse and shift}
The lapse $N$ and the shift $N^k$, which are Lagrange multipliers for the (first-class) constraints,  must be chosen so that the dynamical evolution preserves the boundary conditions.  This means that they can be taken to parametrize a generic asymptotic symmetry.  It is customary to take:
\begin{equation}
	\label{eq:asymptgravII}
	N = 1 + O(r^{-1}), \quad N^r = O(r^{-1}), \quad N^A = O(r^{-2}).
\end{equation}
This corresponds to slicings by hypersurfaces that become asymptotically parallel hyperplanes. Imposing these boundary conditions on the lapse and the shift means that we have to add to the action  the ADM energy, i.e, 
\begin{equation}
B_{S_2} = \oint d^2x
	\sqrt{\xbar \gamma} \,2 \xbar h_{rr}
\end{equation}
(see below).

\subsection{Symplectic form}

We now start to check that the boundary conditions provide a consistent Hamiltonian description.  We first verify that the symplectic structure is well defined.

The leading term of the kinetic term in the action,
\begin{equation}
	\int d^3x \, \pi^{ij} \dot g_{ij} = \int \frac{
		dr}{r} \int d\theta d\phi \, \Big( 2(\xbar
		\pi^{rr} - \xbar \pi^A_A) \dot{\xbar \lambda} + 2\xbar \pi^{AB} 
\dot{\xbar k}_{AB}\Big) + \ldots,
\end{equation}
is generically logarithmically divergent with the decay prescribed by (\ref{eq:asymptgravI}).  The extra conditions are added to cancel this divergence.  

One way to eliminate the logarithmic term would be to take the fields and the corresponding conjugate momenta to have definite and opposite parities so that the integral over the angles vanishes.  There are only two ways to assign these definite parities in a Lorentz invariant manner compatible with the Schwarzschild solution, which are respectively described in \cite{Regge:1974zd} (untwisted case) and \cite{Henneaux:2018cst} (twisted case).  But neither is fully satisfactory (non invariance under the $BMS_4$ algebra, or non-inclusion of the Taub-NUT metric).  

To get a satisfactory phase space where the asymptotic symmetry is the full $BMS_4$ algebra and where the Taub-NUT metric is included, one cannot take the asymptotic fields to have definite parities.  One must allow both untwisted and twisted parity components.  But the logarithmic term in the symplectic form must remain zero.  This is why the twisted parity component is forced to take the specific form of (\ref{eq:parityuptogI})-(\ref{eq:parityuptogII}).  Using integrations by part and the asymptotic constraints \eqref{eq:asymptconst}, one then easily shows that the integral on the sphere appearing in the divergent term of the symplectic structure is indeed always zero.

\subsection{Asymptotic symmetries}

The asymptotic symmetries preserving the boundary conditions are generated by the vector fields
\begin{gather}
	\label{eq:asymptsymI}
	\xi =  b\Big(r - \xbar \lambda -\xbar k\Big) + T + O(r^{-1}),\quad \xi^A =
	Y^A + \frac{1}{r} \Big(\xbar D^A W + \frac{2b}{\sqrt{\xbar\gamma}}
	\xbar \pi^{rA}\Big) + O(r^{-2}),\\
	\label{eq:asymptsymII}
	\xi^r = W + O(r^{-1}), \quad \xbar D_A \xbar D_B b + \xbar \gamma_{AB} b = 0, \quad \cL_Y \xbar
	\gamma_{AB} = 0,
\end{gather}
where $b(x^B), Y^A(x^B)$ describe boosts and spatial rotations, while $T(x^B)$ and $W(x^B)$ are field-independent functions on the sphere.  The choices $T(x^B) \sim Y^{0}_{0} $ and $W(x^B) \sim Y^1_m$ correspond to spacetime translations, but higher spherical harmonics are allowed and involve the $BMS_4$ supertranslations.  

The action of these symmetries on the asymptotic fields is given by
\begin{flalign}
	\delta_\xi \xbar k_{AB} & = \cL_Y \xbar k_{AB} + \xbar D_A \xbar D_B W + W
\xbar \gamma_{AB}\nonumber \\ &\qquad + \frac{b}{\sqrt{\xbar\gamma}} (\xbar \pi_{AB} - \xbar
\gamma_{AB} \xbar \pi^C_C) + \frac{1}{\sqrt{\xbar\gamma}} \xbar D_A( b \xbar
\pi^{rC} \xbar\gamma_{CB}) + \frac{1}{\sqrt{\xbar\gamma}} \xbar D_B( b \xbar
\pi^{rC} \xbar\gamma_{CA}),\\
	\delta_\xi \xbar\lambda &= \frac{b}{4\sqrt 
	{\xbar\gamma}} 
	\xbar p +  Y^C
	\d_C \xbar \lambda,\\
	\delta_\xi(\xbar \pi^{rr} - \xbar \pi^A_A) & = \cL_Y (\xbar \pi^{rr} -
	\xbar \pi^A_A)+\sqrt{ \xbar \gamma} \left( 2 b\xbar
	D_C \xbar D^C \xbar \lambda + 2 \xbar D^C b \d_C \xbar \lambda
+ 6 b \xbar \lambda\right) \\
	\delta_\xi \xbar \pi^{rA} & = \cL_Y \xbar\pi^{rA} + \sqrt{\xbar
	\gamma} \left(\xbar D_B(b \xbar k^{BA}) + \xbar D^A b \xbar k
	- \xbar D^A T\right), \\
	\delta_\xi \xbar \pi^{AB} & = \cL_Y \xbar \pi^{AB} +
	\sqrt{\xbar\gamma} \left( \xbar D^A \xbar D^B T - \xbar \gamma^{AB}
		\xbar D_C \xbar D^C T\right) + 3b \sqrt{\xbar\gamma}\left(\xbar k^{AB} - \xbar
	\gamma^{AB} \xbar k\right) \nonumber \\ & \qquad + \sqrt{\xbar\gamma} b
	\left(\xbar\gamma^{AB} \xbar D_C \xbar D^C \xbar k + \xbar D_C
		\xbar D^C \xbar k^{AB} - \xbar D_C \xbar D^A \xbar
	k^{CB} - \xbar D_C \xbar D^B \xbar k^{CA}\right)\nonumber\\ &
	\qquad +\sqrt{\xbar \gamma} \Big( - \xbar D^A b \xbar D^B
	\xbar k - \xbar D^B b \xbar D^A \xbar k+ \xbar \gamma^{AB} \xbar D_C
	b \xbar D^C \xbar k  + 2 \xbar \gamma^{AB} \xbar D^D \xbar k^C_D \d_C
	b\nonumber\\ & \qquad \qquad - \xbar D^A \xbar k^{BC} \d_Cb-
\xbar D^B \xbar k^{AC} \d_Cb + \xbar D^C \xbar k^{AB} \d_C b\Big).
\end{flalign}

These transformation rules display one important feature.  One can read from them that the variation of the functions $(U)^{odd}$ and $(V)^{even}$ take the following form:
\begin{flalign}
	\delta_\xi (U)^{odd} & = Y^C \d_C (U)^{odd} - b (V)^{even} 
			+ (W)^{odd},\\
	\delta_\xi (V)^{even} & = Y^C \d_C (V)^{even} - 3 b
			(U)^{odd} - \d_A b \xbar D^A (U)^{odd} - b \xbar D_A
			\xbar D^A (U)^{odd} + (T)^{even}.
\end{flalign}
From these, one sees that the variation of these functions under finite supertranslations is additive.  Therefore, the twisted piece in the parity conditions is just that induced by a finite transformation of the asymptotic fields.  It follows that if one starts from a configuration that satisfies the untwisted parity conditions, one generically generates a nonvanishing twist that takes exactly the prescribed form, except if one restricts the transformation to the Poincar\'e algebra in which case the twist remains zero.  Invariance of the boundary conditions under the extended set of transformations is in that sense direct.  That the enhancement, described here by the two functions $T(x^B)$ and $W(x^B)$ on the sphere,  leads exactly to the $BMS_4$ algebra requires further analysis  since the $BMS_4$ supertranslations are characterized by a single function on the sphere.  This was shown in \cite{Henneaux:2018cst}. To understand this point necessitates the form of the charges and is explained in the next section.

\subsection{Charge-generators}

The construction of the charges follows standard Hamiltonian lines  \cite{Regge:1974zd}.  The  steps and difficulties parallel those of the treatment of the twisted case given in \cite{Henneaux:2018cst}, so we only give the result.   Assuming that the asymptotic parameters $T, W, Y^A$ and $b$ are field independent, one finds that the asymptotic symmetries are canonical transformations generated by
\begin{equation}
	\label{eq:bms4gravgen}
	P^{grav}_{\xi}[g_{ij}, \pi^{ij}] = \int d^3x \, \left(\xi \mathcal H + \xi^i \mathcal H_i
	\right) + \mathcal B^{grav}_\xi[g_{ij}, \pi^{ij}].
\end{equation}
Here, as in \cite{Henneaux:2018cst},  the boundary term $\mathcal B_\xi$ is finite thanks to the constraint conditions on the asymptotic fields.  Explicitly, one finds for $\mathcal B_\xi$
\begin{multline}
	\label{eq:boundtermBgen}
	\mathcal B_\xi[g_{ij}, \pi^{ij}] = \oint d^2x \Big
	\{  
	Y^A \Big(4\xbar k_{AB} \xbar \pi^{rB} - 4 \xbar \lambda\xbar \gamma_{AB}
	\xbar\pi^{rB}+ 2 \xbar \gamma_{AB}
	\pi^{(2)rB}\Big) +2 W \Big( \xbar \pi^{rr} -  \xbar \pi^A_A\Big) \\ + T\, 4\sqrt{\xbar\gamma}\,  \xbar
	\lambda +b\,
	\sqrt{\xbar \gamma} \Big( 2  k^{(2)} + \xbar k^2 +
	\xbar k^A_B \xbar k^B_A - 6 \xbar\lambda\xbar k\Big) +b\frac{2}{\sqrt{\xbar\gamma}} \xbar \gamma_{AB}
	\xbar\pi^{rA}\xbar\pi^{rB}\Big\}.
\end{multline}
Note that the charges involve contributions that are quadratic in the asymptotic fields.  These are absent for untwisted parity conditions  \cite{Regge:1974zd}.  By making a BMS transformation away from an ``untwisted frame'', one therefore generates quadratic contributions.

Due to the parity conditions on $\xbar \lambda$ and $\xbar \pi^{rr} - \xbar \pi^A_A$, the transformations generated by even $W$'s and odd $T$'s have zero charge and are proper gauge transformations.  They do not change the physical state of the system and can be factored out.  By contrast, the transformations generated by odd $W$'s and even $T$'s generically have non-vanishing charges.  Such transformations are improper gauge transformations that do change the physical state of the system \cite{Benguria:1976in}.  The physically relevant functions of the angles appearing in the transformations are then $(T)^{even}$ and $(W)^{odd}$.

The algebra is easily evaluated to be:
\begin{equation}
	\Big\{P^{grav}_{\xi_1}[g_{ij}, \pi^{ij}], P^{grav}_{\xi_2}[g_{ij}, \pi^{ij}]\Big\} =
	P^{grav}_{\hat\xi}[g_{ij}, \pi^{ij}],
\end{equation}
where $\hat \xi$ generates an asymptotic symmetry with the following parameters
\begin{flalign}
	\label{eq:hamilbmsI}
	\hat Y^A & = Y^B_1\d_B Y_2^A + \xbar \gamma^{AB} b_1\d_B b_2 - (1
	\leftrightarrow 2),\\
	\label{eq:hamilbmsII}
	\hat b & = Y^B_1\d_B b_2 - (1 \leftrightarrow 2),\\
	\label{eq:hamilbmsIII}
	\hat T & = Y_1^A\d_A T_2 - 3 b_1 W_2 - \d_A b_1 \xbar D^A W_2 - b_1
	\xbar D_A\xbar D^A W_2 - (1 \leftrightarrow 2),\\
	\label{eq:hamilbmsIV}
	\hat W & = Y_1^A \d_A W_2 - b_1T_2 - (1 \leftrightarrow 2).
\end{flalign}
Modding out the trivial transformations generated by even $W$'s  and odd $T$'s, the resulting algebra is the algebra found in \cite{Henneaux:2018cst}. Using the results of \cite{Troessaert:2017jcm}, this algebra was shown there to be the $BMS_4$ algebra expressed in an unfamiliar parametrization. This was done by integrating the equations of motion for the symmetry parameters all the way to null infinity, along the lines explained in the introduction.  One finds that the odd $W$'s  and even $T$'s combine to yield the arbitrary function of the angles parametrizing supertranslations in the original parametrization.  This enables one to conclude that the symmetry at spatial infinity is the same $BMS_4$ as the $BMS_4$ uncovered at null infinity.

In this context, we note that the parity conditions are conditions relating fields at antipodal points on the same asymptotic spheres at spatial infinity. When passing to hyperbolic coordinates to go from spatial infinity to future null infinity and past null infinity, one finds that the parity conditions relate then antipodal points on different spheres since the antipodal transformation is accompanied by a change of sign of the hyperbolic time $\tau$ \cite{Henneaux:2018cst,Troessaert:2017jcm}.   This implies that the values of the fields on asymptotic spheres at null infinity are not restricted by parity conditions.  One gets instead matching conditions  between the values of the fields on the future and past critical spheres (i.e., the past boundary of future null infinity and the future boundary of past null infinity), which involves the antipodal map in agreement with \cite{Strominger:2017zoo}.  One way to get an idea of why the future critical sphere is related to the past critical sphere under parity is to boost arbitrary data fulfilling the parity conditions given on some Cauchy hypersurface.  For each value of the velocity, one gets a new Cauchy hypersurface bounded by an asymptotic sphere on which the parity conditions are satisfied since these are Lorentz invariant.  In the limit of infinite velocity, the parity conditions are still fulfilled, but the asymptotic sphere $S$ meets the critical spheres at points that are antipodally related, so that the antipodal map sends the intersection of $S$ with the future critical sphere on the intersection of $S$ with the past critical sphere.

We also note that the boundary conditions imply that the components of the Weyl tensor do fulfill the definite parity conditions of \cite{Regge:1974zd} since the twisted part drops from them.  For that reason, they remain finite as one goes to null infinity \cite{Troessaert:2017jcm}.  The unwanted feature of generic twisted parity contributions is absent.

\section{Einstein-Maxwell system}
\label{sec:EM}

In this section, we  combine the pure gravity case presented above with results on electromagnetism described in \cite{Henneaux:2018gfi} in order to obtain the asymptotic structure of the Einstein-Maxwell system. We  begin with a short review of the asymptotic structure of Maxwell's theory in Minkowski space and then couple it to gravity in the second subsection.

\subsection{Maxwell field on Minkowski background}
\label{sec:Maxwell}

We start with a review of \cite{Henneaux:2018gfi} to which we refer for details.  We use the same notation as in the previous section in order to describe the background Minkowski metric:
\begin{equation}
	ds^2 = -dt^2 + dr^2 + r^2 \xbar \gamma_{AB} dx^A dx^B.
\end{equation}
The global action of electromagnetism in Minkowski space can then be written as follows:
\begin{multline}
	S_H[A_i, \pi^i, \Psi, \pi_\Psi; A_t, \chi] = \int dt \left\{ \int d^3x \, 
		\pi^i \d_t
		A_i + \pi_\Psi \d_t \Psi - \oint d^2x
		\, \sqrt {\xbar\gamma}\,  \xbar A_r \d_t \xbar \Psi
		\right.\\
		\left. 	- \int d^3x \left(\frac{1}{2\sqrt g} \pi^i \pi_i + \frac{\sqrt
		g}{4} F^{ij} F_{ij} \right)  - \int d^3x \left( \chi \pi_\Psi +  A_t \mathcal{G}  \right)\right\},
\end{multline}
with the following asymptotic behaviour for the fields
\begin{gather}
	\label{eq:falloffsEMI}
	A_r = \frac{1}{r} \xbar A_r + \frac{1}{r^2} A^{(1)}_r +
	o(r^{-2}),\quad
	\pi^r = \xbar \pi^r + \frac{1}{r} \pi^{(1)r} +
	o(r^{-1}),\\
	A_A = \xbar A_A + \frac{1}{r} A^{(1)}_A +
	o(r^{-1}),\quad
	\pi^A = \frac{1}{r}\xbar \pi^A + \frac{1}{r^2} \pi^{(1)A} +
	o(r^{-2}),\\
	\label{eq:elecFalloffIII}
	A_t =  \xbar A_t + \frac{1}{r} A^{(1)}_t +
	o(r^{-2}), \quad \d_i \pi^i = O(r^{-2}),\\
	\label{eq:elecFalloffIV}
	\Psi = \frac 1 r \xbar \Psi + \frac 1 {r^2} \Psi^{(1)} + O(r^{-3}),
	\quad\pi_\Psi =  \frac{1}{r} \pi^{(1)}_\Psi +
	o(r^{-1}),\\
	\chi =\frac 1 r \xbar \chi + \frac 1 {r^2} \chi^{(1)} +
	O(r^{-3}).
\end{gather}
On top of the usual dynamical fields, i.e. the vector potential $A_i$ and the electric field $\pi^i$, we have an extra canonical pair $(\Psi, \pi_\Psi)$.  The fields $A_t$ and $\chi$ are lagrange multipliers for the two first class constraints: Gauss's law $\mathcal G = -\d_i \pi^i\approx 0$ and a new constraint $\pi_\Psi\approx 0$. Locally, this new constraint can be easily solved and one recovers the usual bulk hamiltonian action for electromagnetism. However, it was shown in \cite{Henneaux:2018gfi} that these extra fields and the boundary contribution to the kinetic term are needed for the global description of the system and its symmetries. 

As in the gravity case, one has to impose parity conditions in order to have a finite kinetic term. Various propositions were made in \cite{Henneaux:1999ct,Henneaux:2018gfi}.   We adopt here the ones that allow for the simultaneous description of all know solutions and of the angle dependent $u(1)$-symmetries introduced at null infinity. They can be expressed as \cite{Henneaux:2018gfi}
\begin{equation}
	\label{eq:emParity}
	\xbar A_r = (\xbar A_r)^{odd}, \quad \xbar \pi^r = (\xbar
	\pi^r)^{even}, \quad
	\xbar A_A = (\xbar A_A)^{even} + \d_A \xbar \Phi, \quad \xbar \pi^A = (\xbar
	\pi^A)^{odd},
\end{equation}
where the function $\xbar\Phi$ can be restricted to be an even function on the sphere. One easily checks that the radial logarithmic divergence appearing in the bulk integral of the kinetic term disappears by combining these parity conditions with the asymptotic constraints imposed in \eqref{eq:elecFalloffIII} and \eqref{eq:elecFalloffIV} \cite{Henneaux:2018gfi}.

A particularity of the action is that the symplectic structure $\Omega$ derived from its kinetic term has a boundary term:
\begin{equation}
	\Omega = \int d^3x\Big( d_V
		\pi^i \, d_V A_i + d_V
	\pi_\Psi \, d_V \Psi \Big) - \oint d^2x \, \sqrt{\xbar \gamma}\,
	d_V \xbar A_r \,  d_V
	\xbar \Psi,
\end{equation}
where we used $d_V$ to denote the exterior derivative in phase-space. For that
reason, the prescription $-i_X\Omega = d_V
F$ that associates a function $F$ to a canonical transformation described by the
phase-space vector field $X$ can lead to a surface contribution even if the bulk part of $F$ contains no spatial derivative. 

The action is invariant under two linearly independent gauge transformations: usual electromagnetic gauge transformations and arbitrary shifts of $\Psi$.  Their action on the dynamical fields are
\begin{equation}
	\delta_{\epsilon,\mu} A_i = \d_i \epsilon, \quad 
	\delta_{\epsilon,\mu} \pi^i = 0, \quad 
	\delta_{\epsilon,\mu} \Psi = \mu, \quad 
	\delta_{\epsilon,\mu} \pi_\Psi = 0,
\end{equation}
with the following asymptotic behaviour for the gauge parameters
\begin{equation}
	\epsilon = \xbar \epsilon + \frac 1 r \epsilon^{(1)} + O(r^{-2}),
	\quad \mu = \frac 1 r \xbar \mu + \frac 1 {r^2} \mu^{(1)} + O(r^{-3}).
\end{equation}
Assuming that $\xbar \mu$ and $\xbar \epsilon$ are field independent, their total generator is given by
\begin{equation}
	\label{eq:generatorU1largeI}
	G_{\epsilon, \mu} = \int d^3x(\mu \pi_\Psi + \epsilon \mathcal G) +
	\oint d^2x (\xbar \epsilon\, \xbar \pi^r - \sqrt{\xbar\gamma}\, \xbar \mu
	\xbar A_r).
\end{equation}
Due to the parity conditions \eqref{eq:emParity}, only the transformations for which $\xbar\epsilon$ is even or $\xbar\mu$ is odd are improper gauge transformations. 

The system is also invariant under Poincar\'e transformations. The associated total generator can be written as
\begin{gather}
	\label{eq:PoincEM26}
	P_{\xi, \xi^i} = \int d^3x \left( \xi \mathcal H^{EM} + \xi^i
	\mathcal H^{EM}_i\right) + \mathcal B^{EM}_{(\xi, \xi^i)},\\
	\mathcal H^{EM} = -\Psi \d_i\pi^i -
	 A_i \nabla^i\pi_\Psi + 
	 \frac{1}{2\sqrt g} \pi_i \pi^i +
	 \frac{\sqrt g}{4} F_{ij} F^{ij},\\
	  \quad \mathcal H^{EM}_i = F_{ij} \pi^j - A_i \d_j \pi^j+ \pi_\Psi
	 \d_i \Psi,\\
	\mathcal B^{EM}_{\xi, \xi^i} = \oint d^2x \left(b (\xbar \Psi \xbar \pi^r
		 +\sqrt {\xbar\gamma}  \xbar
		A_A \xbar D^A A_r) +  Y^A(\xbar A_A\xbar \pi^r +
		\sqrt{\xbar\gamma} \,\xbar \Psi
	\d_A\xbar A_r)\right),
\end{gather}
where the Killing vectors of the background $(\xi, \xi^i)$ are given by
\begin{gather}
	\label{eq:asymptsymemI}
	\xi =  br + T,\quad \xi^A =
	Y^A + \frac{1}{r} \xbar D^A W, \quad \xi^r = W,\\
	\label{eq:asymptsymemII}
	\xbar D_A \xbar D_B b + \xbar \gamma_{AB} b = 0, \quad \cL_Y \xbar
	\gamma_{AB} = 0.
\end{gather}
The algebra of the various generators is easily computed:
\begin{gather}
	\label{eq:EMalgebraI}
	\{P_{\xi_1, \xi^i_1},P_{\xi_2, \xi^i_2}\} = P_{\hat\xi, \hat
	\xi^i}, \\
	\{G_{\mu,\epsilon}, P_{\xi,\xi^i}\} = G_{\hat \mu, \hat
	\epsilon},\quad
	\{G_{\mu_1,\epsilon_1}, G_{\mu_2,\epsilon_2}\} = 0,\\
	\hat \xi = \xi_1^i \d_i \xi_2 - \xi_2^i \d_i \xi_1,
	\label{eq:EMalgebraII}
	\quad \hat \xi^i = \xi_1^j \d_j \xi_2^i - \xi_2^j \d_j \xi_1^i +
	g^{ij}(\xi_1 \d_j \xi_2 - \xi_1 \d_j\xi_2),\\
	\label{eq:EMalgebraIII}
	\hat \mu = \nabla^i(\xi\d_i \epsilon)-\xi^i\d_i \mu,\quad
	\hat\epsilon = \xi \mu-\xi^i\d_i \epsilon. 
\end{gather}
The algebra of the symmetries is thus a semi-direct sum of the Poincar\'e algebra and the abelian algebra parametrized by $\xbar\mu$ and $\xbar\epsilon$, the action of the Poincar\'e subalgebra characterising this semi-direct sum being given by:
\begin{gather}
	\delta_{(Y,b,T,W)}\xbar \mu = Y^A\d_A \xbar \mu - \xbar D_A(b \xbar D^A\xbar
	\epsilon), \quad \delta_{(Y,b,T,W)}\xbar \epsilon = Y^A \d_A \xbar
	\epsilon - b \xbar \mu.  \label{eq:EMalgebraXX}
\end{gather}
It was also shown in \cite{Henneaux:2018gfi} that this algebra agrees with the one obtained at null infinity. The core idea is that the even and odd functions, $\xbar \epsilon$ and $\xbar \mu$, combine to form a single function on the sphere that generates the angle-dependent $u(1)$-transformations seen at null infinity.

\subsection{Combining gravity and electromagnetism}
\label{sec:einsteinmaxell}

Combining all the results described in the previous sections for gravity and electromagnetism is straightforward.   The starting point is the following action for the Einstein-Maxwell system:
\begin{flalign}
	\label{eq:hamiltactionfull}
	S & = \int dt \left\{\int d^3x \Big(\pi^{ij} \d_t g_{ij} + \pi^i \d_t A_i
		+ \pi_\Psi \d_t \Psi\Big) - \oint d^2x \sqrt {\xbar\gamma} \, \xbar A_r \d_t
		\xbar \Psi \right.\nonumber\\
		&\left. \qquad - \int d^3x \Big(\chi
	\pi_\Psi + A_t \mathcal{G} +N^i \mathcal H_i + N \mathcal H\Big)  - \oint d^2x \sqrt{\xbar
\gamma} \, 2 \xbar h_{rr}\right\},\\
	\mathcal H & = -\sqrt g R + \frac{1}{\sqrt g} (\pi^{ij} \pi_{ij} -
	\frac{1}{2} \pi^2)  - \Psi \partial_i \pi^i - A_i \nabla^i \pi_\Psi + \frac{\sqrt g}{4} F_{ij} F^{ij} + \frac{1}{2}
	\frac{1}{\sqrt g} \pi^i \pi_i,\\
	\mathcal H_i & = -2 \nabla_j \pi^j_i + F_{ij} \pi^j - A_i \d_j \pi^j +
	\pi_\Psi \d_i A_t.
\end{flalign}
The asymptotic conditions appropriate to this action are the ones we described previously for gravity and electromagnetism: see equations \eqref{eq:asymptgravI} to \eqref{eq:asymptgravII}  for the gravitational field and equations \eqref{eq:falloffsEMI} to \eqref{eq:emParity} for the electromagnetic one.

We assume in particular that the constraints hold asymptotically, in the sense that they fall off at least one order faster than the one implied by the boundary conditions on the fields,
\begin{equation}
	\mathcal H = o(r^{-1}), \quad
	\mathcal H_r = o(r^{-1}), \quad
	\mathcal H_A = o(1), \quad
	\mathcal \d_i \pi^i = o(r^{-1}).
\end{equation}
As the contribution of the electromagnetic field to the gravitational constraints is sub-leading, the asymptotic conditions on the gravitational fields are unchanged
\be 
\label{eq:EinstMaxAsympconst}
\xbar \pi^{rA} +\xbar D_B \xbar
\pi^{BA} = 0, \quad \xbar D_A \xbar D_B \xbar \pi^{AB} + \xbar \pi^A_A = 0,
\quad \xbar D_A \xbar D_B \xbar k^{AB} -
	\xbar D_A \xbar D^A \xbar k = 0.
\ee

The generators of the large $u(1)$ gauge transformations  in equation \eqref{eq:generatorU1largeI} are easily seen to remain allowed functionals in the combined case. To build the generators of the $BMS_4$ transformations, we  add the gravitational generators $P_\xi^{grav}$ of equation \eqref{eq:bms4gravgen} to the Poincar\'e generators \eqref{eq:PoincEM26} of electromagnetism:
\begin{equation}
	P_\xi = P^{grav}_\xi + P^{EM}_\xi,
\end{equation}
and allow the diffeomorphism generators $\xi = (\xi^\perp, \xi^i)$ to take the same general form as in the pure gravitational case:
\begin{gather}
	\xi^\perp =  b\Big(r - \xbar \lambda - \xbar k\Big) + T + O(r^{-1}),\quad \xi^A =
	Y^A + \frac{1}{r} \Big(\xbar D^A W + \frac{2b}{\sqrt{\xbar\gamma}}
\xbar \pi^{rA}\Big) + O(r^{-2}),\\
	\xi^r = W + O(r^{-1}), \quad \xbar D_A \xbar D_B b + \xbar \gamma_{AB} b = 0, \quad \cL_Y \xbar
	\gamma_{AB} = 0.
\end{gather}
This gives the expression
\begin{equation}
	\label{eq:fullbms4fullaction}
	P_\xi  =  \int d^3x \, \left(\xi^\perp \mathcal H + \xi^i \mathcal
	H_i\right) + \mathcal B_\xi,
\end{equation}
where the boundary term reads
\begin{multline}
	\mathcal B_\xi[g_{ij}, \pi^{ij}] = \oint d^2x \Big
	\{ T\, 4\sqrt{\xbar\gamma}\,  \xbar
	\lambda + 2 W \Big( \xbar \pi^{rr} -  \xbar \pi^A_A\Big)  \\
	+Y^A \Big(4\xbar k_{AB} \xbar \pi^{rB} - 4 \xbar \lambda\xbar \gamma_{AB}
	\xbar\pi^{rB}+ 2 \xbar \gamma_{AB}
	\pi^{(2)rB} +\xbar A_A\xbar \pi^r + \sqrt{\xbar\gamma} \,\xbar \Psi
	\d_A\xbar A_r\Big) + \\ + b\,
	\sqrt{\xbar \gamma} \Big( 2  k^{(2)} + \xbar k^2 +
	\xbar k^A_B \xbar k^B_A - 6 \xbar\lambda\xbar k + \xbar \Psi \xbar \pi^r
		 +\sqrt {\xbar\gamma}  \xbar
		A_A \xbar D^A A_r\Big) +b\frac{2}{\sqrt{\xbar\gamma}} \xbar \gamma_{AB}
	\xbar\pi^{rA}\xbar\pi^{rB}\Big\}.
\end{multline}
Using the previous results, we can easily check that
\begin{equation}
	d_V P_\xi = -i_\xi \Omega,
\end{equation}
where $\Omega$ is the symplectic structure of the Einstein-Maxwell action given in \eqref{eq:hamiltactionfull}:
\begin{equation}
	\Omega = \int d^3x\Big( d_V \pi^{ij} \, d_V g_{ij} + d_V
		\pi^i \, d_V A_i + d_V
	\pi_\Psi \, d_V \Psi \Big) - \oint d^2x \, \sqrt{\xbar \gamma}\,
	d_V \xbar A_r \,  d_V
	\xbar \Psi.
\end{equation}
As in the pure gravitational case, the a priori divergent term produced by the variation of the generator disappears using the asymptotic constraints \eqref{eq:EinstMaxAsympconst}.  Moreover, one can also check that the associated variations preserve the asymptotic conditions on the canonical variables. The two properties together prove that the generators written in \eqref{eq:fullbms4fullaction} are allowed functionals. 

The algebra of the constraints leads to the following algebra for the gauge parameters:
\begin{equation}
	[(\xi_1^\perp, \xi_1^i, \mu_1, \epsilon_1), 
	(\xi_2^\perp, \xi_2^i,\mu_2, \epsilon_2)]_M = 
	(\hat\xi^\perp, \hat\xi^i, \hat\mu, \hat\epsilon),
\end{equation}
where
\begin{gather}
	\hat\xi^i  \approx [\xi_1,\xi_2]_{SD}^i+
	\delta_2 \xi_1^i - \delta_1 \xi_2^i, \qquad
	\hat\xi^\perp  \approx [\xi_1,\xi_2]_{SD}^\perp +
	\delta_2 \xi_1^\perp - \delta_1 \xi_2^\perp, \\
	\hat\mu  \approx -\nabla^i(\xi_1^\perp \d_i\epsilon_2)+\xi_1^i\d_i\mu_2
	+ A_t \nabla^i(\xi_1^\perp\d_i \xi_2^\perp) 
	- \frac{\pi}{2\sqrt g} A^i\xi_1^\perp\d_i \xi_2^\perp
	+\delta_2 \mu_1- (1
	\leftrightarrow 2), \\
	\hat \epsilon  \approx -\xi_1^\perp \mu_2 +\xi_1^i\d_i\epsilon_2 +
	\delta_2 \epsilon_1 - (1
	\leftrightarrow 2).
\end{gather}
The variations $\delta_1$ and $\delta_2$ denotes the action of the gauge transformations acting on the canonical fields. From this, we can read off the algebra of the asymptotic gauge parameters using the transformation laws of the asymptotic fields:
\begin{gather}
	\label{eq:fullhamilu1bmsi}
	\hat Y^A  = Y^B_1\d_B Y_2^A + \xbar \gamma^{AB} b_1\d_B b_2 - (1
	\leftrightarrow 2),\quad
	\hat b  = Y^B_1\d_B b_2 - (1 \leftrightarrow 2),\\
	\label{eq:fullhamilu1bmsii}
	\hat T  = Y_1^A\d_A T_2 - 3 b_1 W_2 - \d_A b_1 \xbar D^A W_2 - b_1
	\xbar D_A\xbar D^A W_2 - (1 \leftrightarrow 2),\\
	\label{eq:fullhamilu1bmsiv}
	\hat W  = Y_1^A \d_A W_2 - b_1T_2 - (1 \leftrightarrow 2),\\
	\label{eq:fullhamilu1bmsv}
	\hat \mu  =  Y_1^A\d_A \xbar \mu_2 +\xbar D_A(b_1
	\xbar D^A\xbar
	\epsilon_2)- (1 \leftrightarrow 2),\quad
	\hat \epsilon  = Y_1^A \d_A \xbar
	\epsilon_2 + b_1 \xbar \mu_2- (1 \leftrightarrow 2).
\end{gather}
As before, the transformations generated by $W$ and $\xbar \mu$ even and $T$ and $\xbar\epsilon$ odd are proper gauge transformations and have to be modded out.

The total algebra of asymptotic symmetries $\mathcal A$ is therefore the semi-direct sum of Lorentz algebra with the direct sum of the two abelian algebras representing supertranslations and large $u(1)$ gauge transformations:
\begin{equation}
	\mathcal A = \text{Lorentz} \oplus_\sigma (\text{Supertranslations} \oplus
	u(1)\text{-transformations}).
\end{equation}
According to general theorems \cite{Brown:1986ed}, the algebra of the charges  reproduces this algebra, here without central extension:
\begin{equation}
	\big\{ G(\xi_1^\perp, \xi_1^i, \mu_1, \epsilon_1), 
	G(\xi_2^\perp, \xi_2^i, \mu_2, \epsilon_2)\big\} = 
	G(\hat\xi^\perp, \hat\xi^i, \hat\mu, \hat\epsilon).
\end{equation}
The simplest way to check the absence of central extension is to evaluate the Poisson bracket on the vacuum: Minkowski space with zero electromagnetic field.  This algebra is the globally well-defined one obtained at null infinity in previous analyses \cite{Barnich:2013sxa}.

\section{Conclusions}
\label{sec:Conclusions}

In this note, we have provided a description of the Einstein-Maxwell system at spatial infinity that possesses the following necessary features:
\begin{enumerate}
	\item  Solutions with both gravitational electric and gravitational magnetic mass belong to the phase-space defined by the boundary conditions.  Electromagnetic magnetic monopoles are also included.
	\item The symplectic form, and thus the kinetic term in the action, is finite, so that the canonical structure is well-defined.
	\item The asymptotic symmetry algebra is the $BMS_4$ algebra, or the semi-direct sum of the $BMS_4$ algebra with the abelian algebra of angle-dependent $u(1)$ gauge transformations when electromagnetism is included.  All elements of that algebra have  well-defined canonical generators (otherwise, they would not be true symmetries).
\end{enumerate}
The $BMS_4$ algebra is here the same $BMS_4$ algebra uncovered at null infinity.   In that sense it is a bit misleading to talk about asymptotic symmetries ``at spatial infinity'' since the terms ``at spatial infinity''  are superfluous.

The boundary conditions at spatial infinity involve in an essential way parity conditions on the leading orders of the fields in their asymptotic expansion. These parity conditions are Lorentz-invariant and acceptable since they do not exclude known physically interesting solutions.  They match the analysis at null infinity, not only by yielding the same symmetries, but also by leading to fields on the critical spheres bounding future null infinity and past null infinity that fulfill the matching conditions adopted there \cite{Strominger:2017zoo}.  The parity conditions are necessary to make the symplectic form finite.  In perfect analogy with the Maxwell case \cite{Henneaux:2018gfi} the boundary conditions can be obtained by acting with a general supertranslation on the set of solutions satisfying the original untwisted parity conditions of \cite{Regge:1974zd}.  They take a particularly simple form, because the integrated variations of the relevant functions take the same form as the infinitesimal variations (abelian action).

One motivation behind our work was to provide a direct Hamiltonian description of the infrared structure and of the charges underlying the soft graviton/photon theorems \cite{Ashtekar:1981bq,Ashtekar:1981sf,Ashtekar:1987tt,Strominger:2013jfa,He:2014laa,Cachazo:2014fwa,Strominger:2014pwa,Pasterski:2015tva,Campiglia:2015kxa,Conde:2016rom,He:2014cra,Lysov:2014csa,He:2015zea,Kapec:2015ena,Campiglia:2016hvg,Conde:2016csj,Campiglia:2017mua,Laddha:2017vfh}. That one can achieve such a description is not surprising, not only because everything is anchored on Cauchy hypersurfaces, but also because the matching conditions underlying the soft theorems are intimately related to the ``Coulomb behaviour'' of the  fields \cite{Strominger:2013jfa,Strominger:2017zoo} (and of the dual  ``magnetic Coulomb behaviour'' when there are magnetic charges).  This behaviour is of course recorded at spatial infinity, and we have given the action of the $BMS_4$ group on the asymptotic fields there.  Even though there is no gravitational or electromagnetic radiation reaching spatial infinity, the action of the group is non trivial on these asymptotic data.  In particular, the asymptotic fields corresponding to the Minkowski solution ($\xbar k_{AB} = 0 = \xbar \lambda = \xbar \pi^{rr} = \xbar \pi^{rA} = \xbar \pi^{AB}$) do transform under supertranslations.   The Minkowski solution is not invariant and belongs to a non trivial orbit of the action of the $BMS_4$ algebra, although its charges remain zero. 

As one goes to null infinity, the electromagnetic field  or the Weyl tensor  generically develops a logarithmic singularity for twisted parity conditions, \cite{Troessaert:2017jcm}, \cite{Henneaux:2018gfi} (see also \cite{Herberthson:1992gcz}).  This problem is part of the general question on how Cauchy data behave as one approaches null infinity  (see \cite{Christodoulou:1993uv,Bieri:2009xc,Friedrich:2017cjg,Hintz:2017xxu,Paetz:2018nbd} and references therein for the relevant literature).  Cauchy data without logarithmic singularities may diverge logarithmically in that null limit, which, more generally, needs to be described by a polyhomogeneous expansion in $r^i \log^j r$.    By taking the twist to be trivial, i.e., given by a gauge transformation that is allowed to be improper, we eliminate the first, divergent piece in that expansion.  Stronger smoothness conditions at null infinity would imply stronger restrictions on the Cauchy data,  but these would still have to satisfy the boundary conditions adopted here in order to fulfill the leading regularity condition at null infinity.

Finally, we note that our boundary conditions lead to the original $BMS_4$ algebra as asymptotic symmetry algebra,  without super-rotations \cite{Banks:2003vp,Barnich:2010eb,Barnich:2009se}.  This question deserves further study.


\section*{Acknowledgments} 
Fruitful discussions with the participants in the Solvay workshop ``Infrared Physics: Asymptotic \& BMS symmetry, soft theorems, memory, information paradox and all that'' (Brussels, 16 - 18 May 2018) are gratefully acknowledged. This work was partially supported by the ERC Advanced Grant ``High-Spin-Grav" and by FNRS-Belgium (convention FRFC PDR T.1025.14 and  convention IISN 4.4503.15).

\begin{appendix}

\section{More details on the parity conditions}
\label{app}
\subsection{Tensor decomposition}

The coefficient of the divergent term present in the gravitational action has the form
\begin{equation}
	\label{eq:surfdiverg}
	\oint d^2x \, \left( 2 (\xbar \pi^{rr} - \xbar \pi^A_A) \dot{\xbar
	\lambda} + 2 \xbar \pi^{AB} \dot {\xbar k}_{AB}\right).
\end{equation}
In order for this term to be zero, we need the various tensors to belong to orthogonal subspaces. There is little choice for the first term as we want the solution space to contain black-holes and be invariant under Poincar\'e transformations, we must impose the parity conditions described in \cite{Regge:1974zd}. On the other hand, there is more freedom in the choice of conditions on the angular part.

In order to better understand the asymptotic structure, it is interesting to introduce a decomposition of rank 2 symmetric tensors on the sphere that is adapted to the gravitational fields. A systematic decomposition of tensors on the sphere can be done using spin coefficients. However, as we only have rank 1 and 2 tensors in this work, it will be more convenient to use a decomposition based on scalar potentials.  

On the sphere, a 1-form $V_A$ can be decomposed into a transverse part $V_A^T$ and a longitudinal part $V_A^L$ as follows 
\begin{equation}
	V_A = V_A^L + V_A^T, \quad V_A^L = \d_A V^L, \quad V^T_A =  e_A^{\phantom AB}
	\d_B V^T,
\end{equation}
where $e_{AB}$ is the anti-symmetric tensor with $e_{\theta\phi} = \sin \theta$ in usual spherical coordinates. The two scalar functions $V^L$ and $V^T$ are potentials and are defined up to a constant. The decomposition is orthogonal in the sense that:
\begin{equation}
	\oint d^2x \sqrt {\xbar\gamma} \,\xbar \gamma^{AB} \,V_A W_B = \oint d^2x
	\sqrt {\xbar\gamma} \,\xbar \gamma^{AB} \left( V^T_A W^T_B + V^L_AW^L_B \right).
\end{equation}

A similar decomposition can be introduced for rank 2 symmetric tensors $T_{AB}$. Indeed, for all such tensors, there exists a vector $V_A$ and a scalar $\hat T$ such that
\begin{equation}
	T_{AB} = \xbar D_A V_B + \xbar D_B V_A + \xbar\gamma_{AB} \hat T.
\end{equation}
This property is the linearised version of the statement that all metrics on the sphere are conformally equivalent: any deformation of the metric, here $\delta\xbar\gamma_{AB} = T_{AB}$, is the sum of an infinitesimal diffeomorphism and an infinitesimal Weyl transformation. We then introduce the decomposition of the vector $V_A$ described above to obtain
\begin{equation}
	T_{AB} = 2\xbar D_A\xbar D_B V^L + \left(e_A^{\phantom AC} \xbar D_B \xbar D_C
	V^T+e_B^{\phantom AC} \xbar D_A \xbar D_C V^T\right) + \xbar \gamma_{AB} \hat T.
\end{equation}
This decomposition is valid in general and is equivalent to the one given in term of spin coefficients\footnote{Introducing stereographic coordinates on the sphere such that $\xbar \gamma_{AB} dx^A dx^B = 2 P^{-2} d\zeta d\xbar \zeta$ and the covariant derivative is encoded in the $\eth$ and $\xbar\eth$ operators, we see that the three combinations $\xbar\eth^2 V^L -i \xbar\eth^2 V^T$, $\eth^2 V^L +i \eth^2 V^T$ and $P^2(\hat T + \Delta V^L)$ correspond to the three spin coefficients of weights 2, -2 and 0 describing a symmetric rank 2 tensor.}. However, it is not well adapted to the gravitational problem and needs to be slightly modified by introducing alternative potentials. For all pairs of functions $(V^L, \hat T)$, there exists a unique pair $(T^L, T^T)$ such that
\begin{equation}
	2 V^L = T^L + T^T, \qquad \hat T = T^L -\Delta T^T.
\end{equation}
This can be easily proved using the fact that the operator $\Delta + 1$ is invertible on the sphere. Renaming the potential $V^T$ as $T^{TT}$, we proved that a symmetric rank 2 tensor $T_{AB}$ can be uniquely decomposed into three parts $T^L_{AB}$, $T^T_{AB}$ and $T^{TT}_{AB}$ as follows
\begin{equation}
	T_{AB} = T^{TT}_{AB} + T^T_{AB} + T^L_{AB},
\end{equation}
where
\begin{flalign}
	T^{TT}_{AB} & = e_A^{\phantom AC} \xbar D_B \xbar D_C
	T^{TT}+e_B^{\phantom AC} \xbar D_A \xbar D_C T^{TT}, \\
	T^T_{AB} & =
	\xbar D_A \xbar D_B T^T - \xbar \gamma_{AB} \Delta T^T, \\
	T^L_{AB} &= \xbar D_A \xbar D_B T^L + \xbar \gamma_{AB} T^L.
\end{flalign}
The names of the various parts ($TT$, $T$ and $L$) are related to the way they appear in the gravitational constraints and are inspired by the decomposition of the bulk 3 dimensional rank 2 tensors. As in the rank 1 case, the decomposition is orthogonal:
\begin{equation}
	\oint d^2x \sqrt{\xbar \gamma} \,\xbar \gamma^{AC} \xbar \gamma^{BD}
	\, T_{AB} U_{CD} = 
	\oint d^2x \sqrt{\xbar \gamma} \,\xbar \gamma^{AC} \xbar \gamma^{BD}
	\left( T^{TT}_{AB} U^{TT}_{CD} + T^T_{AB} U^T_{CD} +T^L_{AB} U^L_{CD} \right).
\end{equation}

Introducing this decomposition for $\xbar k_{AB}$ and $\xbar \pi^{AB}$, one easily shows that the asymptotic constraints \eqref{eq:asymptconst} impose the following conditions
\begin{flalign}
	\label{eq:appAsymptConstk}
	\xbar D^A \xbar D^B \xbar k_{AB} - \Delta \xbar k = 0 & \Rightarrow \qquad
	\xbar k^T_{AB} = 0,\\
	\xbar D_A \xbar D_B \xbar \pi^{AB} + \xbar \pi^A_A = 0 & \Rightarrow
	\qquad \xbar\pi^{LAB} = 0,
\end{flalign}
where the decomposition is extended to densities in the natural way.  The surface integral \eqref{eq:surfdiverg} controlling the logarithmic divergence of the action then simplifies to
\begin{equation}
	\oint d^2x \,\xbar \pi^{AB} \xbar k_{AB} = 
	\oint d^2x  \,
	\xbar \pi^{TTAB} \xbar k^{TT}_{CD}.
\end{equation}
The only components on which we have to impose parity conditions are these TT components while the "pure gauge" components $\xbar\pi^{TAB}$ and $\xbar k^L_{AB}$ can be left arbitrary. The choice presented in section \ref{sec:grav} corresponds to $\xbar k^{TT}_{AB}$ even and $\xbar \pi^{TTAB}$ odd while the choice leading to the twisted parity conditions of \cite{Henneaux:2018cst} corresponds to the opposite. In the next part of this appendix, we will show how the $TT$ components of both tensors are related to the asymptotic behaviour of the Weyl tensor.

\subsection{Weyl tensor}
\label{app:weyl}

The electric and magnetic parts of the 4 dimensional Weyl tensor $W_{\mu\nu\rho\sigma}$ are given in terms of canonical fields by the following expressions
\begin{equation}
	\mathcal E_{ij} = R_{ij} + \frac 1 g \left( \frac 1 2 \pi \pi_{ij} -
		\pi_{ik} \pi^k_j\right), \qquad \mathcal B_{ij} = - \frac 2
		{\sqrt g} e_j^{\phantom j kl} \nabla_k \left( \pi_{il} - \frac
		1 2  \pi g_{il}\right),
\end{equation}
where $e_{ijk}$ is the totally antisymmetric tensor such that $e_{r \theta \phi} = \sqrt g$. In asymptotically euclidean coordinates, for which the metric and the momenta take the following form
\begin{equation}
	g_{ij} = \delta_{ij} + O(r^{-1}), \qquad \pi^{ij} = O(r^{-2}),
\end{equation}
both tensors have the following behaviour
\begin{equation}
	\mathcal E_{ij} = r^{-3} \xbar{\mathcal E}_{ij} +  O(r^{-4}), \qquad \mathcal
	B_{ij} = r^{-3} \xbar{\mathcal B}_{ij} + O(r^{-4}).
\end{equation}

Using spherical coordinates, and the asymptotic conditions of section \ref{sec:grav}, their leading term takes the form
\begin{flalign}
	\mathcal E_{AB} &= \frac 1 r \left(\xbar D_C \xbar D_B \xbar
	k^C_A + \xbar D_C \xbar D_A \xbar k^C_B  - \xbar D_C \xbar D^C \xbar
	k_{AB} - \xbar D_A \xbar D_B \xbar k - 2 \xbar k_{AB} + \xbar k \xbar
	\gamma_{AB}\right)\nonumber \\ 
	& \qquad + \frac 1 r \left( \xbar \gamma_{AB} \xbar D_C
	\xbar D^C \xbar \lambda - \xbar D_A \xbar D_B \xbar \lambda + \xbar
\gamma_{AB} \xbar \lambda \right) + O(r^{-2}),\\
\mathcal E_{rA} & = \frac 1 {r^2} \left( \d_A \xbar k - \xbar D_B \xbar
k^B_A\right) + O(r^{-3}),\\
\mathcal E_{rr} & = \frac 1 {r^3} \left(- \xbar D_A\xbar D^A \xbar \lambda - 2
\xbar \lambda\right) + O(r^{-4}),\\
\mathcal B_{AB} &= \frac 1 r \, \frac{-1}{\sqrt{\xbar\gamma}}\left(e_{AC}
\xbar\gamma_{BD}
(\xbar \pi^{CD} + \xbar D^C \xbar \pi^{rD}) + e_{BC}
\xbar\gamma_{AD}
(\xbar \pi^{CD} + \xbar D^C \xbar \pi^{rD})\right) + O(r^{-2}),\\
\mathcal B_{rA} & = \frac 1 {r^2} \,\frac {-1} {\sqrt{\xbar \gamma}}\left(
	e_{AC}( \xbar D^C \xbar \pi^{rr} + \xbar \pi^{rC}) + 
e^{CD} \xbar D_C \xbar\pi_{DA}\right) + O(r^{-3}),\\
\mathcal B_{rr} & = \frac 1 {r^3} \,\frac 2 {\sqrt{\xbar \gamma}}
	e^{\phantom AB}_A
\xbar D_B \xbar \pi^{rA} + O(r^{-4}).
\end{flalign}
One easily checks that these asymptotic expressions only depend on supertranslation invariant quantities
\begin{equation}
	\xbar \lambda, \quad \xbar \pi^{rr} - \xbar \pi^A_A, \quad \xbar
	k^{TT}_{AB}, \quad \xbar \pi^{TTAB}.
\end{equation}
As such, the new parity conditions introduced in section \ref{sec:grav} lead to the following parity behaviour for the asymptotic electric and magnetic Weyl tensors:
\begin{equation}
	\label{eq:parityWeyl}
	\xbar {\mathcal E}_{AB} \sim \xbar {\mathcal E}_{rr} \sim \xbar
	{\mathcal B}_{AB} \sim
	\xbar {\mathcal B}_{rr} = \text{even}, \quad \xbar {\mathcal E}_{Ar} 
	\sim \xbar {\mathcal B}_{rA} = \text{odd}.
\end{equation}
or, equivalently, in Cartesian coordinates:
\begin{equation}
	\xbar{\mathcal E}_{ij} \sim \xbar{\mathcal B}_{ij} = \text{even}.
\end{equation}

The converse is also true in the sense that, if the parity conditions are imposed on the radial components $\xbar\lambda$ and $\xbar\pi^{rr} - \xbar\pi^A_A$ then imposing the parity conditions \eqref{eq:parityWeyl} on the Weyl tensor is equivalent to imposing the standard untwisted parity conditions on the $TT$ part of the angular components $\xbar k^{TT}_{AB}$ and $\xbar\pi^{TTAB}$:
\begin{equation}
	\xbar k^{TT}_{AB} = \text{even}, \quad \xbar \pi^{TTAB} = \text{odd},
\end{equation}
which is equivalent to the relaxed parity conditions given in
\eqref{eq:parityuptogI}-\eqref{eq:parityuptogII}.

\end{appendix}

\end{document}